\documentclass[conference]{IEEEtran}
\IEEEoverridecommandlockouts

\usepackage{cite}

\ifCLASSINFOpdf
\else
\fi

\usepackage{amsthm}

\usepackage{amsmath,amssymb,mathtools,dsfont,bm}

\usepackage{color}

\usepackage{algorithm,algorithmic}

\DeclarePairedDelimiter{\norm}{\lVert}{\rVert}

\usepackage{subcaption}

\makeatletter
\renewenvironment{thebibliography}[1]{%
	\@xp\section\@xp*\@xp{\refname}%
	\normalfont\footnotesize\labelsep .5em\relax
	\renewcommand\theenumiv{\arabic{enumiv}}\let\p@enumiv\@empty
	\vspace*{-5pt}
	\list{\@biblabel{\theenumiv}}{\settowidth\labelwidth{\@biblabel{#1}}%
		\leftmargin\labelwidth \advance\leftmargin\labelsep
		\usecounter{enumiv}}%
	\sloppy \clubpenalty\@M \widowpenalty\clubpenalty
	\sfcode`\.=\@m
}{%
	\def\@noitemerr{\@latex@warning{Empty `thebibliography' environment}}%
	\endlist
}
\makeatother

\let\OLDthebibliography\thebibliography
\renewcommand\thebibliography[1]{
	\OLDthebibliography{#1}
	\setlength{\parskip}{-.4pt}
	\setlength{\itemsep}{-0.7pt}
}
\IEEEsettopmargin{t}{56pt}
\IEEEoverridecommandlockouts\IEEEpubid{\makebox[\columnwidth]{ 978-1-6654-5975-
		4/22~\copyright~2022 IEEE \hfill} \hspace{\columnsep}\makebox[\columnwidth]{ }}
\begin{document}
	%
	\title{Joint Communication and Computation in Hybrid Cloud/Mobile Edge Computing Networks}
	
	
	\author{\IEEEauthorblockN{Robert-Jeron Reifert\IEEEauthorrefmark{1},
			Hayssam Dahrouj\IEEEauthorrefmark{2},
			Basem Shihada\IEEEauthorrefmark{3},
			Aydin Sezgin\IEEEauthorrefmark{1},\\
			Tareq Y. Al-Naffouri\IEEEauthorrefmark{3} and
			Mohamed-Slim Alouini\IEEEauthorrefmark{3}}
		\IEEEauthorblockA{\IEEEauthorrefmark{1}Institute of Digital Communication Systems, Ruhr-Universit\"at Bochum, Germany
		}
		\IEEEauthorblockA{\IEEEauthorrefmark{2}Department of Electrical Engineering, University of Sharjah, Sharjah, United Arab Emirates}
		\IEEEauthorblockA{\IEEEauthorrefmark{2}Communication Theory Lab, King Abdullah University of Science and Technology, Thuwal, Saudi Arabia}}
	
	
	\maketitle
	\footnotetext[1]{This work was was supported in part by the German Federal Ministry of Education and Research (BMBF) in the course of the 6GEM Research Hub under Grant 16KISK037.}

\begin{abstract}
	Facing a vast amount of connections, huge performance demands, and the need for reliable connectivity, the sixth generation of communication networks (6G) is envisioned to implement disruptive technologies that jointly spur connectivity, performance, and reliability. In this context, this paper proposes, and evaluates the benefit of, a hybrid central cloud (CC) computing and mobile edge computing (MEC) platform, especially introduced to balance the network resources required for joint computation and communication. Consider a hybrid cloud and MEC system, where several power-hungry multi-antenna unmanned aerial vehicles (UAVs) are deployed at the cell-edge to boost the CC connectivity and relieve part of its computation burden. While the multi-antenna base stations are connected to the cloud via capacity-limited fronthaul links, the UAVs serve the cell-edge users with limited power and computational capabilities. The paper then considers the problem of maximizing the weighted network sum-rate subject to per-user delay, computational capacity, and power constraints, so as to determine the beamforming vectors and computation allocations. Such intricate non-convex optimization problem is tackled using an iterative algorithm that relies on $\ell_0$-norm relaxation, successive convex approximation, and fractional programming, and has the compelling ability to be implemented in a distributed fashion across the multiple UAVs and the CC.	The paper results illustrate the numerical prospects of the proposed algorithm for enabling joint communication and computation, and highlight the appreciable improvements of data processing delays and throughputs as compared to conventional system strategies.
\end{abstract}


%
\IEEEpeerreviewmaketitle

\section{Introduction}
Today's Internet of Things (IoT) applications involve many relevant consumer and industry use cases, e.g., smart cities, modular plant, etc. For the next generation of wireless communication systems, massive IoT is forecasted to make up of $51$\% cellular IoT connections by $2027$, while other use-cases extend to augmented reality, vehicular to anything, etc. \cite{emr}. With massively increased number of connected devices and increased service requirements, massive IoT raises further challenges towards the realization of the sixth generation of communication networks (6G).
Due to their finite energy and computation resources, IoT devices are often dependent on offloading their tasks, especially for computation-intensive applications \cite{8961914}.
A suitable technique for satisfying such massive data demand is the cloud-based network architecture, which enables centralized management of communication and computation resources. However, a drawback of cloud-based networks is the long propagation delay \cite{8016573} and the need for costly, limited-capacity fronthaul links to connect the cloud to the base stations (BSs).
Moving computation and management capabilities towards the network edge enables both latency reduction and service quality enhancement. Such cost-efficient and energy-saving paradigm, referred to as mobile edge computing (MEC), is subject, however, to strict constraints on power and computational resources.
{To this end, this paper considers one particular hybrid network architecture, where the central cloud connects to central BSs so as to serve the central network users. The cell-edge users, on the other hand, are served by resource-limited MEC devices, especially deployed to boost the system connectivity. The paper then adopts such a hybrid cloud/MEC architecture to empower joint communication and computation by means of maximizing a network-wide sum-rate, the performance of which is a function of the allocated computation and communication resources. } 

{The topic considered in this paper is related to the general context of resource management in cloud-radio access networks \cite{Journal,DaiY14}, and MEC-based works focusing on computation \cite{8809879,8434285,8961914} and communication \cite{8016573,8757041,8580994}.}
Unmanned aerial vehicles (UAV)-assisted MEC has been recognized as a promising 6G network technique for allowing flexible deployment, on-demand service, and enhanced connectivity \cite{8016573,8764580,9461747}. The utilization of UAVs, with typically strict power constraints, calls for sophisticated joint management of communication and computation resources in order to optimize the system performance.
Due to weak received power and strong adjacent network interference, especially in the prospective 6G ultra-dense deployment, cell-edge users are often prone to inferior service quality, which makes the development of communication and computation resource management techniques vital. 
Related works in this field include UAV-aided communication \cite{8757041,8580994} and computation \cite{8809879,8445936} networks. These works, however, do not capture the joint consideration of the communication and computation aspects, i.e., constraint-wise and variable-wise, e.g., see \cite{8764580}.
As 6G communication networks are envisioned to include multiple access technologies in a hybrid manner, the need to consider the interplay of a central network and edge devices arises. In contrast to related UAV-focused literature, e.g., \cite{8764580,9461747,9615116}, this work extends the joint communication and computation paradigm toward hybrid cloud and MEC networks.
Further, in contrast to previous works on MEC networks which adopt orthogonal access schemes, e.g., \cite{8952621,8877759}, this paper adopts a spatial multiplexing approach and separates users using a beamforming strategy. Reference \cite{Journal} utilizes a similar approach for resource management under the multi-cloud paradigm; however, the essential computation and delay considerations are ignored in \cite{Journal} which rather focuses on mitigating the intra- and inter-cloud interference in a multi-cloud setup in the absence of any MEC capabilities.

{Unlike the aforementioned references, this paper proposes a downlink hybrid cloud/MEC network, where several multi-antenna BSs and UAVs serve single-antenna network users. The BSs are connected to the cloud via capacity limited fronthaul links, while the UAVs perform computation and communication functions on their own.}
We address a sum-rate maximization problem by jointly managing beamforming vectors, allocated rates, and computation capacity, subject to per-BS and per-UAV power, per-BS fronthaul capacity, per-computing platform maximum computation capacity, and per-user delay constraints. Such mixed discrete-continuous non-convex optimization problem is tackled using $\ell_0$-norm relaxation, successive convex approximation (SCA), and fractional programming (FP) resulting in a fully centralized protocol (FCP) and in an efficient partially decentralized protocol (PDP). Insightful simulations verify the gains of the proposed network architecture in terms of sum-rate and delay. The proposed decentralized algorithm is particularly shown to overcome the centralized version in terms of computational complexity, runtime, and scalability, as well as the fully distributed protocol (FDP) in terms of sum-rate.

\section{System Model and Problem Formulation}\label{sec:sysmod}
\begin{figure}[!t]
\centering
\includegraphics[width=3.2in]{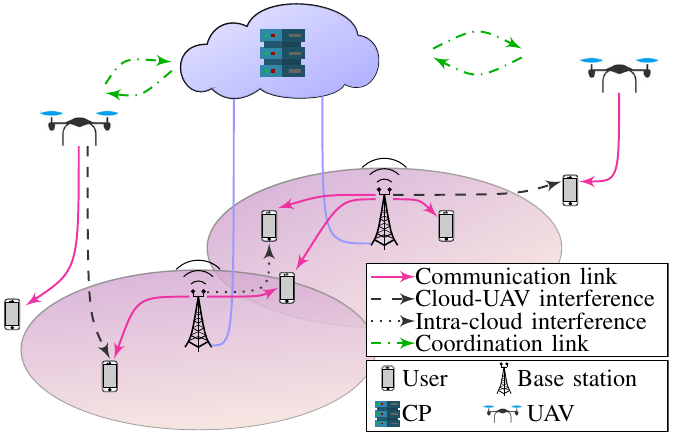}
\vspace*{-.2cm}
\caption{Network of $2$ UAVs, $2$ BSs, and $6$ users.}
\label{sys_mdl}
\vspace*{-0.8cm}
\end{figure}

In this work, we consider the downlink of a hybrid cloud/MEC-based network architecture. Under such framework, cloud processors (CPs) coordinate the users operations within the core-network. The UAVs, on the other hand, with on-chip computation capabilities, act as mobile edge computers (ECs) to serve the cell-edge users.
The core network consists of a single cloud, i.e., the central cloud (CC), connected via fronthaul links to $B$ multi-antenna BSs, with $L_\text{c}$ antennas each, while the UAVs at the edge are equipped with $L_\text{e}$ antennas each. 
The split of network functions follows a \emph{data-sharing} approach, where the CP at the CC performs most network functions, e.g., encoding and precoder design, leaving the modulation, precoding, and radio tasks to the BSs \cite{DaiY14}. 
Fig.~\ref{sys_mdl} shows an example of the considered system, which illustrates a network of $2$ BSs serving $4$ central users, and $2$ UAVs each serving one user. 
Let $E$ be the number of deployed ECs, and let $\mathcal{E}=\{1,\cdots,E\}$ be the set of ECs. Since each EC is implemented on a UAV, the edge network consists of $E$ UAVs. 
Note that throughout this work, the terms ECs and UAVs are interchangeably used without loss of generality. The set of BSs is given by $\mathcal{B}=\{1,\cdots,B\}$.
The set of single-antenna users is denoted by $\mathcal{K}=\left\{1,\cdots,K\right\}$, where $K$ is the total number of users. In the context of CC and EC coexistence, this paper assumes disjoint user-clusters, which are covered by the user sets {$\mathcal{K}_\text{c}\subseteq\mathcal{K}$ and $\mathcal{K}_e\subseteq\mathcal{K}$}, with $\mathcal{K}_\text{c} \cap \mathcal{K}_e  = \emptyset, \forall e\in\mathcal{E}$ and $\mathcal{{K}}_e \cap \mathcal{{K}}_e'  = \emptyset, \forall e\neq e'$. In other terms, the set of users served by the CC is denoted by $\mathcal{K}_\text{c}$, while the set of users served by each EC $e$ is denoted by $\mathcal{K}_e$, $\forall e\in\mathcal{E}$. Similarly, the CC serves $K_c$ users, while EC $e$ serves $K_e$ users. The determination of the sets $\mathcal{K}_e$ and $\mathcal{K}_c$ falls outside the scope of the current paper, as it is often determined on a different time-scale than the beamforming problem considered in this paper.

We further denote the channel vector from BS $b$ to user $k$ by $\mathbf{h}_{b,k}\in\mathbb{C}^{L_\text{c}}$, and the channel vector from UAV $e$ to user $k$ by $\tilde{\mathbf{h}}_{e,k}\in\mathbb{C}^{L_\text{e}}$, where $e\in\mathcal{E}$. The aggregate channel vector from all BSs and UAVs towards user $k$ is given by $\mathbf{h}_{k} = [\mathbf{h}_{1,k}^T,\cdots,\mathbf{h}_{B,k}^T,\tilde{\mathbf{h}}_{1,k},\cdots,\tilde{\mathbf{h}}_{E,k}]^T$. For mathematical tractability, the paper assumes full knowledge of channel state information at the transmitters (CSIT).\\	
\indent The paper then aims at jointly managing the communication and computation network resources, which are captured through the following list of variables:\vspace{-.1cm}
\begin{itemize}
\item The \emph{beamforming vectors} $\mathbf{w}_{b,k}\in\mathbb{C}^{L_\text{c}}$ and $\tilde{\mathbf{w}}_{e,k}\in\mathbb{C}^{L_\text{e}}$, which denote the beamforming vector of user $k$'s signal at BS $b$ and at UAV $e\in\mathcal{E}$, respectively.
\item The \emph{computation vector} $\mathbf{f}\in\mathbb{N}^K$, which denotes the allocated computation cycles to process the users data, where each entry $f_k$ is given in cycles/s, $\forall k \in \mathcal{K}$.
\item The \emph{rate allocation vector} $\mathbf{r}\in\mathbb{R}^K$, which denotes the achievable rates of all users, with $\mathbf{r} = [r_1,\cdots,r_K]$.\vspace{-.1cm}
\end{itemize}
Note that $\mathbf{w}_{k} = [\mathbf{w}_{1,k}^T,\cdots,\mathbf{w}_{B,k}^T,\tilde{\mathbf{w}}_{1,k}^T,\cdots,\tilde{\mathbf{w}}_{E,k}^T]^T$ is the aggregate beamforming vector for user $k$. As per the user association constraints, $\mathbf{w}_{k}$ is group-sparse by design since $k$ may only be served by either CC or one EC. Additionally, in the case when the CC serves the user $k$, not all BSs participate in serving $k$, i.e., $\mathbf{w}_{b,k} = \mathbf{0}^{L_\text{c}}$ for some BSs $b$. We next describe the expressions of the metrics relevant to the paper context, mainly, the data-rates, the power consumption at the BSs and at the UAVs, the users transmission delays, and the computation capacity.\vspace{-.05cm}
\paragraph{Achievable Rate}
Each user receives its own intended signal $\mathbf{h}_{k}^{\dagger} \mathbf{w}_{k} s_k$, and treats all other users' signals as interference. This is captured in the signal to interference plus noise ratio (SINR) of user $k$ as\vspace*{-.2cm}
\begin{equation}
\Gamma_{k} = \frac{|\mathbf{h}_{k}^{\dagger} \mathbf{w}_{k}|^2}{\sigma^2 + \sum_{i\in\mathcal{K}\backslash\{k\}} |\mathbf{h}_{k}^{\dagger} \mathbf{w}_{i}|^2}, \label{SINR}\vspace{-.2cm}
\end{equation}
where $\sigma^2$ is the additive white Gaussian noise variance. The achievable rate of user $k$ is then expressed as\vspace*{-.2cm}
\begin{equation}
r_k \leq \tau \mathrm{log}_2(1+\Gamma_{k}),\vspace*{-.2cm}
\end{equation}
where $\tau$ is the transmission bandwidth. Keeping all signals to be transmitted by a BS $b$ within the limits of the finite fronthaul capacity 
$R_b^\text{max}$, we write the fronthaul capacity constraint as\vspace*{-.15cm}
\begin{equation}
\sideset{}{_{k\in\mathcal{K}}}\sum \norm[\big]{\norm[\big]{\mathbf{{w}}_{b,k}}_{2}^{2}}_0 r_{k} \leq R_b^\text{max}, \qquad\forall b \in \mathcal{B}, \label{eq:front}\vspace*{-.1cm}
\end{equation}
where the $\ell_0$-norm in \eqref{eq:front} is a non-linear function that determines the number of non-zero elements. Hence, we determine if a BS $b$ serves user $k$ solely based on the beamforming vector, i.e., if $b$ does not assign any power to serve $k$, the rate of $k$ does not contribute to BS $b$'s fronthaul link.
\paragraph{Power Consumption}
This work distinguishes between two power consumption metrics: $P_{b}^\mathrm{cc}$, i.e., the power consumed by BS $b$, and $P_{e}^\mathrm{ec}$, i.e., the power consumption of EC $e$, which are defined respectively as:\vspace*{-.2cm}
\begingroup	\addtolength{\jot}{-.1cm}
\begin{align}
&P_{b}^\mathrm{cc} = \sideset{}{_{k\in\mathcal{K}_\text{c}}}\sum \norm[\big]{\mathbf{{w}}_{b,k}}_{2}^{2}, \label{eq:pcc} \\
&P_{e}^\mathrm{ec} = \underbrace{\sideset{}{_{k\in\mathcal{K}_e}}\sum\norm[\big]{\tilde{\mathbf{w}}_{e,k}}_{2}^{2}}_{\text{Transmission}} + \underbrace{s_e \left(\sideset{}{_{k\in\mathcal{K}_e}}\sum f_{k} \right)^{\mu_e}}_{\text{Computation}} +\underbrace{\phantom{\Big(}Q_e\phantom{\Big)}}_{\text{Operation}}\hspace{-.1cm}, \label{eq:pec}
\end{align}\endgroup
\\\vspace*{-.8cm}\\
where $s_e$ and $\mu_e$ are constants that depend on the CPU model \cite{8434285}. The expressions in \eqref{eq:pcc} and \eqref{eq:pec} denote that while the BSs are assumed to have a maximum communication transmit power constraint, the mobile ECs are subject to more elaborate power constraints. Especially, we assume each EC to have a maximum power consumption constraint, that consists of transmit, computation, and operational power, see \cite{8434285}. The first two terms in \eqref{eq:pec} are directly coupled with beamforming and computation variables, respectively, while the latter term $Q_e$ is fixed, accounting for flight-related mechanical and operational power. 
\paragraph{Delay Considerations} The overall delay that each user experiences is a function of the following components:\vspace*{-.2cm}
\begin{align}
&\Theta_{k} = \underbrace{{F_k}/{f_{k}}}_{\text{Computation Delay}} + \underbrace{\phantom{/}\Lambda_{k}\phantom{/}}_{\text{Fronthaul Delay}} + \underbrace{{D_k}/{r_k}}_{\text{Transmission Delay}},\label{eq:delaycc}
\end{align}
\\\vspace{-.8cm}\\
\noindent where $F_k$, $\Lambda_{k}$, and $D_k$ denote the total computation cycles required for user $k$'s task, the worst-case fronthaul delay of all BSs serving $k$, and the data size for $k$, respectively. Note that the fronthaul delay depends on the type of fronthaul link in the network.
More specifically, equation \eqref{eq:delaycc} consists of a computation delay, i.e., the time it takes to process $k$'s requested task, a fronthaul delay, i.e., the time-loss on the fronthaul link, and a transmission delay, i.e., the latency during the wireless data transfer. Note that for the delay of user $k$ served by EC $e$, i.e., if $k$ is an edge-user, the fronthaul delay is rather negligible, and is ignored in the context of our paper.
\paragraph{Computation Capacity} In this paper, each computation platform (i.e., CC or EC's) is subject to a specific maximum computation capacity constraint as follows:\vspace*{-.2cm}
\begin{equation}
\sideset{}{_{k\in\mathcal{K}_\text{c}}}\sum f_{k} \leq f_{c}^{\text{max}}; \quad \sideset{}{_{k\in\mathcal{K}_e}}\sum f_{k} \leq f_{e}^{\text{max}}, \; \forall e\in\mathcal{E}, \label{eq:maxcompcap}\vspace*{-.15cm}
\end{equation}
where $f_{c}^{\text{max}}$ and $f_{e}^{\text{max}}$ denote the CC's and EC $e$'s capacity, respectively. 
\subsection{Problem Formulation}
Given the above considerations, the paper then aims at maximizing a weighted sum-rate by managing the beamforming vectors, allocated rates, and computation allocation. 
The mathematical formulation of the considered problem can then be written as follows:\vspace*{-.2cm}
\begin{subequations}\label{eq:Opt1}
\begingroup
\addtolength{\jot}{-.08cm}
\begin{align}
\underset{\mathbf{w},\mathbf{r},\mathbf{f}}{\text{max}}\quad &\sideset{}{_{k\in\mathcal{K}}}\sum \alpha_{k} r_{k}  \tag{\ref{eq:Opt1}} \\
\text{s.t.} \quad\; & \eqref{SINR}, \eqref{eq:front}, \eqref{eq:pec}, \eqref{eq:maxcompcap}, \nonumber\\
&r_{k} \leq \tau\,\text{log}_2(1+\Gamma_{k}),\hspace*{1.0cm} &\forall k \in \mathcal{K}, \label{eq:achr}\\
&\sideset{}{_{k\in\mathcal{K}_\text{c}}}\sum \norm[\big]{\mathbf{{w}}_{b,k}}_{2}^{2} \leq P_{b}^{\text{max}},\hspace*{-5cm} &\forall b \in \mathcal{B}, \label{eq:powercc1}\\
&P_{e}^\mathrm{ec}(\mathbf{w},\mathbf{f}) \leq P_{e}^{\text{max}},\hspace*{-8cm}  &\forall e \in \mathcal{E}, \label{eq:powerec1}\\
&{F_k}/{f_{k}} + {D_k}/{r_k} \leq t_k - \Lambda_{k},\hspace*{-9cm} & \forall k \in \mathcal{K}_\text{c}, \label{eq:delaycc1}\\
&{F_k}/{f_{k}} + {D_k}/{r_k} \leq t_k,\hspace*{-9cm} & \forall k\in\mathcal{K}_e, \forall e\in\mathcal{E}. \label{eq:delayec1}\\[-0.65cm]\nonumber
\end{align}
\endgroup
\end{subequations}
where $\alpha_k$ is the fixed weight associated with $r_k$, $\mathbf{w} = [\mathbf{w}_1^T,\cdots,\mathbf{w}_K^T]^T$, and $\mathbf{f} = [f_1,\cdots,f_K]^T$. Problem \eqref{eq:Opt1}'s feasible set is defined by the fronthaul capacity constraint per BS \eqref{eq:front} (non-convex, mixed-integer), the maximum computation capacity per cloud \eqref{eq:maxcompcap} (convex), the achievable rate per user \eqref{eq:achr} (non-convex, fractional), the maximum transmit power constraint per BS \eqref{eq:powercc1} and per UAV \eqref{eq:powerec1} (convex), the delay constraint per CC's users \eqref{eq:delaycc1} and per ECs' users \eqref{eq:delayec1} (convex), where $t_k$ denotes the maximum tolerable delay per user. Due to the nature of the $\ell_0$-norm, and the non-convexity of the feasible set, problem \eqref{eq:Opt1} is a mixed-integer non-convex optimization problem. As such problems are generally difficult to solve, our paper next proposes a series of problem reformulations so as  to devise a numerically practical algorithm to solve problem \eqref{eq:Opt1}.
\section{Proposed Algorithm}\label{sec:alg}\vspace*{-.1cm}
Given the numerical intricacies of problem \eqref{eq:Opt1}, the paper now proposes a practically feasible algorithm, the highlight of which is its ability of being implemented in a distributed fashion across both the CC and the ECs.
Initially, the mixed-integer nature of constraint \eqref{eq:front} is tackled using a heuristic $\ell_0$-norm approximation. Subsequently, SCA finds a convex upper-bound to tackle the non-convex nature of the relaxed constraint \eqref{eq:front2}. After introducing an auxiliary variable, FP decouples the numerator and denominator in \eqref{SINR}. Each step introduces auxiliary variables, which either contribute to the set of optimization variables, or are updated in an outer loop. We next present each of the above optimization formulations in details.\vspace*{-.1cm}
\subsection{User-to-BS Clustering}\label{UBS}\vspace*{-.1cm}
To tackle the discrete structure of the $\ell_0$-norm in \eqref{eq:front}, we utilize a heuristic $\ell_1$-norm approximation with fixed weights in each iteration as\vspace*{-.2cm}
\begin{equation}
\sideset{}{_{k\in\mathcal{K}}}\sum \beta_{b,k}\norm[\big]{\norm[\big]{\mathbf{{w}}_{b,k}}_{2}^{2}}_1 r_{k} \leq R_b^\text{max}, \qquad\forall b \in \mathcal{B}, \label{eq:front2}\vspace*{-.25cm}
\end{equation}
where the weights are defined as $\beta_{b,k} = \big(\delta + \norm[\big]{\mathbf{{w}}'_{b,k}}_{2}^{2}\big)^{-1}$, with $\delta > 0$ and $\mathbf{w}'_{b,k}$ denotes the beamforming vector from the last iteration \cite{DaiY14}. Due to the inverse relationship between $\beta_{b,k}$ and $\mathbf{w}'_{b,k}$, BS-user-links with low allocated power are likely to be deactivated, as high weights pose a burden on the limited capacity fronthaul link. Naturally, only BS-user-links, with reasonable allocated power, remain active at the end of the algorithm. Such formulation enables the opportunity to introduce an auxiliary variable $q_{b,k} = \beta_{b,k}\norm[\big]{\norm[\big]{\mathbf{{w}}_{b,k}}_{2}^{2}}_1$, so as to obtain a bilinear formulation of \eqref{eq:front2}'s left hand side as $q_{b,k} r_{k}$. Define the aggregate vector $\mathbf{q}$ as $\mathbf{q} = [q_{1,1},\cdots,q_{1,K},q_{2,1},\cdots,q_{B,K}]^T$. To be able to apply the SCA framework, e.g., see \cite{Journal}, write $q_{b,k} r_{k}$ as $q_{b,k} r_{k} \triangleq \frac{1}{4} (\left(q_{b,k}+r_{k}\right)^2 - \left(q_{b,k} - r_{k}\right)^2)$.
Using the SCA approach enables us to replace its concave part, i.e., $- \left(q_{b,k} - r_{k}\right)^2$, using the linear first-order Taylor expansion. The reformulated fronthaul capacity constraint \eqref{eq:front} becomes\vspace*{-.15cm}
\begingroup	\addtolength{\jot}{-.2cm}
\begin{align}
&\sideset{}{_{k\in\mathcal{K}}}\sum\big( \left(q_{b,k}+r_{k}\right)^2 - 2 \left(q'_{b,k}-r'_{k}\right)\left(q_{b,k}-r_{k}\right) \nonumber\\
&\qquad\qquad\qquad + \left(q'_{b,k}-r'_{k}\right)^2\big) \leq 4 R_b^\text{max}, \qquad\forall b \in \mathcal{B} \label{eq:front4}.
\end{align}\endgroup
\\\vspace*{-1.0cm}\\
Here $q'_{b,k}$ and $r'_{k}$ denote feasible fixed values, i.e., the optimal optimization variables of the previous iteration. Similar to $\beta_{b,k}$, $q'_{b,k}$ and $r'_{k}$ are updated after each iteration.
At this point, \eqref{eq:front4} is a convex constraint and problem \eqref{eq:Opt1}'s non-convexity solely stems from the achievable rate constraint \eqref{eq:achr}.\vspace{-.1cm}
\subsection{Auxiliary Variables}\vspace{-.1cm}
The highly coupled and fractional SINR expression \eqref{SINR} within the concave logarithm in constraint \eqref{eq:achr} prevents from finding solutions to problem \eqref{eq:Opt1} efficiently. We, therefore, introduce an auxiliary variable for reformulating the complicated constraint \eqref{eq:achr} as $\bm{\gamma}=[\gamma_{1},\cdots,\gamma_{K}]^T$. The reformulated problem, by including the auxiliary variable $\bm{\gamma}$ and the relaxed fronthaul constraint \eqref{eq:front4}, is mathematically formulated as\vspace*{-.2cm}
\begin{subequations}\label{eq:Opt4}
\begingroup
\addtolength{\jot}{-.08cm}
\begin{align}
\underset{\mathbf{w},\mathbf{r},\mathbf{f},\mathbf{q},\bm{\gamma}}{\text{max}}\quad &\sideset{}{_{k\in\mathcal{K}}}\sum \alpha_{k} r_{k} \tag{\ref{eq:Opt4}}\\
\text{s.t.} \quad\;\;\;\, & \eqref{SINR}, \eqref{eq:pec}, \eqref{eq:maxcompcap}, \eqref{eq:powercc1}-\eqref{eq:delayec1}, \eqref{eq:front4},\hspace*{-1.35cm} \nonumber\\
&r_{k} \leq \tau\,\text{log}_2(1+\gamma_{k}),\hspace*{-.2cm} &\forall k \in \mathcal{K}, \label{eq:achr4}\\
&\gamma_{k} \leq \Gamma_{k},\hspace*{-8cm} &\forall k \in \mathcal{K}, \label{eq:sinr4}\\
&\beta_{b,k}\norm[\big]{\norm[\big]{\mathbf{{w}}_{b,k}}_{2}^{2}}_1 \leq q_{b,k}, \hspace*{-.0cm} &\forall b \in \mathcal{B},\forall k\in\mathcal{K}. \label{eq:q}
\end{align}\endgroup
\end{subequations}
\\\vspace{-1.0cm}\\
Note that both \eqref{eq:achr4} and \eqref{eq:q} are convex constraints, and so constraint \eqref{eq:sinr4} remains the main reason for the non-convexity of problem \eqref{eq:Opt4}, and is therefore tackled next in the text.\vspace{-.2cm}
\subsection{Fractional Programming and Algorithm}\vspace{-.1cm}
Based on the general FP framework developed in \cite[Theorem 2]{FP1}, we now apply the quadratic transform in a multidimensional and complex form to decouple the numerator and the denominator of the fractional constraint \eqref{eq:sinr4} as\vspace*{-.2cm}
\begingroup	\addtolength{\jot}{-.14cm}
\begin{align}
&g_k(\mathbf{w}) = \gamma_{k} - 2 \text{Re}\left\{ u_{k}^\dagger {\mathbf{w}}_{k}^\dagger {\mathbf{h}}_{k} \right\} \nonumber\\
&\qquad\qquad\qquad+ |u_{k}|^2 \left[ \sigma^2 + \sideset{}{_{i\in\mathcal{K}\backslash\{k\}}}\sum |{\mathbf{h}}_{k}^\dagger {\mathbf{w}}_{i}|^2 \right] \label{eq:g1},
\end{align}\endgroup
\\\vspace{-.75cm}\\
where $\mathbf{u} = [u_1,\cdots,u_K]^T$ is a vector of complex-valued auxiliary variables. Note that \eqref{eq:g1} is a convex function of the beamforming vector, in case $\mathbf{u}$ is fixed. The optimal $u_{k}$ for fixed $\mathbf{w}$ can be written as\vspace*{-.2cm}
\begin{equation}
u_{k}^* = {{\mathbf{w}}_{k}^\dagger {\mathbf{h}}_{k} }{\Big[ \sigma^2 + \sideset{}{_{i\in\mathcal{K}\backslash\{k\}}}\sum |{\mathbf{h}}_{k}^\dagger {\mathbf{w}}_{i}|^2 \Big]^{-1}}, \label{eq:uck_opt}\vspace{-.2cm}
\end{equation}
which is obtained by setting the partial derivative of $g_k(\mathbf{w})$ with respect to $u_{k}$ to zero and then solving for $u_{k}$.
Hence, under all previously mentioned reformulations and manipulations, the original problem \eqref{eq:Opt1} is approximated by the following computationally tractable optimization problem
\phantom{bla}\vspace{-.6cm}
\begin{subequations}\label{eq:Opt5}
\begingroup
\addtolength{\jot}{-.08cm}
\begin{align}
\underset{\mathbf{w},\mathbf{r},\mathbf{f},\mathbf{q},\bm{\gamma}}{\text{max}}\quad &\sideset{}{_{k\in\mathcal{K}}}\sum \alpha_{k} r_{k} \tag{\ref{eq:Opt5}}\\
\text{s.t.} \quad\;\;\;\, & \eqref{eq:pec}, \eqref{eq:maxcompcap}, \eqref{eq:powercc1}-\eqref{eq:delayec1}, \eqref{eq:front4}, \eqref{eq:achr4}, \eqref{eq:q},\hspace*{-1.8cm} \nonumber\\
&g_k(\mathbf{w}) \leq 0,\hspace*{-8cm} &\forall k \in \mathcal{K}. \label{eq:sinr5}
\end{align}\endgroup
\end{subequations}
\\\vspace{-1cm}\\
As the objective \eqref{eq:Opt5} is linear and the constraints define a convex set, problem \eqref{eq:Opt5} is a convex optimization problem, which can be solved efficiently \cite{cvx}. More specifically, the solution is now computed in an iterative manner, via alternating between solving problem \eqref{eq:Opt5} and updating the auxiliary variable $\mathbf{u}$, the weights $\beta_{b,k}$, and the feasible fixed values $q'_{b,k}$ and $r'_k$, $\forall b\in\mathcal{B}$, $\forall k\in\mathcal{K}$. The detailed steps of the FCP for resource management are presented in Algorithm \ref{AlgCen} above.
\setlength{\textfloatsep}{0pt}
\begin{algorithm}[t]
\caption{Centralized Protocol for Resource Management}
\begin{algorithmic}[1]
\STATE Initialize $\mathbf{w}$ to feasible values\\ \vspace*{-.05cm}
\textbf{Repeat:} until convergence \vspace*{-.05cm}
\STATE Update $\beta_{b,k}$, $q'_{b,k}$, and $r'_k$ as in \ref{UBS}; update $\mathbf{u}^*$ using \eqref{eq:uck_opt}\vspace*{-.05cm}
\STATE Solve convex optimization problem \eqref{eq:Opt5}\vspace*{-.05cm}
\STATE \textbf{End}  \vspace*{-.05cm}
\end{algorithmic}
\label{AlgCen}
\end{algorithm}%
\paragraph{Complexity of the Fully Centralized Solution}
The above FCP's complexity depends on (a) the convergence rate of Algorithm \ref{AlgCen}, i.e., the maximum number of iterations, and (b) the complexity of problem \eqref{eq:Opt5}. The convex problem \eqref{eq:Opt5} can be solved using an interior-point method, where the total number of variables is $\xi_\text{FCP} = K(3+EL_\text{e}+B(1+L_\text{c}))$ \cite{Lobo1998ApplicationsOS}. Hence, the upper-bound computational complexity of the FCP is $\mathcal{O}({V}_{\text{max}}(\xi_\text{FCP})^{3.5})$, where ${V}_{\text{max}}$ is the worst-case number of iterations. Note that, for fixed clustering, high-quality beamforming vectors and computation capacities can be found with reduced computational complexity.\vspace{-.1cm}
\subsection{Decentralized and Distributed Resource Management}\vspace*{-.15cm}
\paragraph{Partially Decentralized Operation}
The paper now illustrates how the proposed algorithm above is amenable for distributed implementation in the considered hybrid cloud/MEC network, with a reasonable amount of information exchange. Note that initially, the CSIT has to be known at all clouds. This is reasonable, as in practice, the UAVs are employed by the network operator by exchanging specific coordination information during UAV-operation, e.g., see Fig.~\ref{sys_mdl}. The distribution of optimization variables towards their respective management entity, i.e., CC or ECs, becomes then feasible, except for constraint \eqref{eq:sinr5}, which requires the CC and ECs to exchange the following terms iteratively\vspace*{-.2cm}
\begin{equation}
\sideset{}{_{i\in\mathcal{K}\backslash\{k\}}}\sum |{\mathbf{h}}_{k}^\dagger {\mathbf{w}}_{i}|^2. \label{eq:interinfo}\vspace*{-.15cm}
\end{equation}
As we assume full CSIT, CC and ECs are able to efficiently compute (\ref{eq:interinfo}) and then forward a single value to all other entities. A procedure for decentralized resource management builds upon Algorithm \ref{AlgCen}, which is implemented at the CC and each EC and adds an additional step between step $3$ and step $4$: \emph{Exchange \eqref{eq:interinfo} between CC and ECs}. Using the PDP, the computational complexity is substantially reduced and distributed among CC and ECs. The computation complexity at the CC becomes $\mathcal{O}({V}_{\text{max}}^\text{pdp}(\xi_\text{PDP})^{3.5})$, and at each EC $\mathcal{O}({V}_{\text{max}}^\text{pdp}(K_\text{e}(3+L_\text{e}))^{3.5})$, where $\xi_\text{PDP}=K_\text{c}(3+B+BL_\text{c})$ and ${V}_{\text{max}}^\text{pdp}$ is the worst-case number of iterations.
\paragraph{Fully Distributed Operation} For benchmarking purposes, we also apply the FPD, i.e., a implementation which treats all cloud-to-EC interference as background noise. In such case, the entities are unaware of each other's operation and manage their respective resources independently. The complexity is analogous to PDP, with a lower worst-case number of iterations, i.e., ${V}_{\text{max}}^\text{pdp}>{V}_{\text{max}}^\text{fdp}$, as the additional information exchange step is dropped. For the ease of presentation, Tab.~\ref{tb:complexity} summarizes the three protocols' complexities. Note that the considered FDP constitutes a rather optimistic procedure, as allocated rates are set to feasible values after solving the problem at the CC and all ECs.
\begin{table}[t]
\vspace*{.1cm}
\renewcommand{\arraystretch}{1.05}
\centering
\begin{tabular}{c | c | c}
\hline
\hspace*{-.2cm}Protocol\hspace*{-.15cm} & Central cloud & Edge cloud \\\hline\hline
FCP & \multicolumn{2}{c}{\hspace*{-.15cm}$\mathcal{O}({V}_{\text{max}}( K(3+EL_\text{e}+B(1+L_\text{c})))^{3.5})$\hspace*{-.15cm}} \\
PDP & \hspace*{-.15cm}$\mathcal{O}({V}_{\text{max}}^\text{pdp}(K_\text{c}(3+B+BL_\text{c}))^{3.5})$\hspace*{-.15cm} & $\mathcal{O}({V}_{\text{max}}^\text{pdp}(K_\text{e}(3+L_\text{e}))^{3.5})$\\
FDP & \hspace*{-.15cm}$\mathcal{O}({V}_{\text{max}}^\text{fdp}(K_\text{c}(3+B+BL_\text{c}))^{3.5})$\hspace*{-.15cm} & $\mathcal{O}({V}_{\text{max}}^\text{fdp}(K_\text{e}(3+L_\text{e}))^{3.5})$\\
\hline
\end{tabular}
\vspace*{-.15cm}
\caption{Computational complexities.}
\label{tb:complexity}
\vspace*{-.1cm}
\end{table}%
\section{Simulations}\label{sec:sim}
\vspace*{-.1cm}
To assess the numerical performance of the proposed algorithms, we consider a cellular network of $7$ BSs, each equipped with 3 antennas, and with $400$m inter-BS distance. At the network edge, we consider $E$ single-antenna ($L_\text{e}=1$) UAVs, each placed at $150$m altitude. The users are placed randomly within the network, with $E$ users at the cell-edge. For illustration, we set the noise power-spectral density to $-134$dBm/Hz, the fronthaul capacity to $R_b^\text{max} = 50$Mbps, $Q_e = 100$W, $F_k = 10^8$cycles, $D_k = 10^5$bits, $f_0^\text{max}=5\cdot10^{10}$cycles/s, $f_e^\text{max}=10^{9}$cycles/s, $P_b^\text{max} = 24$dBm, $P_e^\text{max} = 17$dBm, $\tau = 10$MHz, $\Lambda_{k}=0.45$ms, $t_k = 600$ms, and $\alpha_{k}=1$, $\forall k\in\mathcal{K}, b\in\mathcal{B}$, and $e\in\mathcal{E}$, unless mentioned otherwise. The beamforming vectors are initialized randomly.
In this work, we utilize the 3GPP specified pathloss $p_{k,n}(dB) = 128.1 + 37.6 \log_{10}(d_{k,n})$, with $d_{k,n}$ being the distance between user and BS in km, and the UAV-user channel pathloss modeled in \cite{pl}, with the excessive pathloss component $\eta_\text{los} = 4$, and $\eta_\text{nlos} = 35$. The adopted channel model consists of pathloss, Rayleigh fading ($\Gamma^l_{k,n}\sim \mathcal{CN}(0,1)$), and log-normal shadowing effect ($8$dB standard deviation).
Similar to \cite{8764580}, we set $s_e = 10^{-28}$ and $\mu_e = 3$, utilizing a specific CPU model.


\subsection{Impact of the Number of UAVs}
\begin{figure}[!t]
\centering
\vspace*{-.3cm}
\hspace*{-.2cm}\includegraphics[width=3.74in]{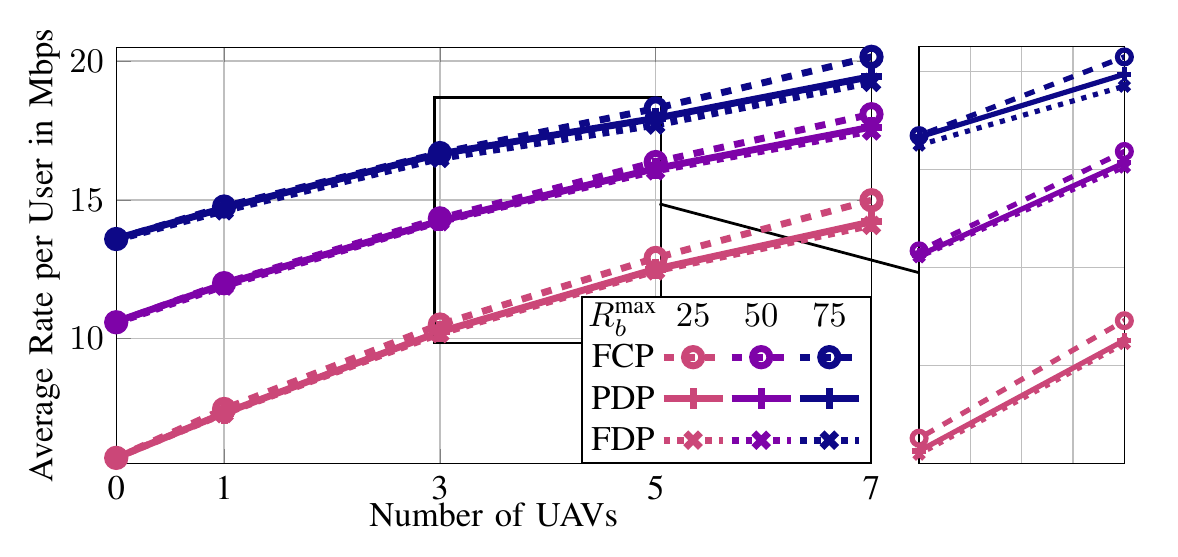}
\vspace*{-.4cm}
\caption{Average rate per user as a function of the number of UAVs for three fronthaul capacities $R_b^\text{max}$ in Mbps.}
\label{yAR_xMEC_F25_F75}
\end{figure}

\vspace*{-.1cm}
In the first set of simulations, we compare FCP, PDP, and FDP in terms of average rate per user versus the number of UAVs for three fronthaul capacities in Fig.~\ref{yAR_xMEC_F25_F75}, with $K=30$ users. 
It is first observed that all three algorithmic implementations of the considered scheme, i.e., centralized, decentralized, and fully distributed, show considerable improvements with increasing number of UAVs. This emphasizes the positive impact of the UAV-aided MEC model adopted in the paper. Interestingly, when $R_b^\text{max} = 25$Mbps, all schemes outperform the $0$ UAVs-case with $R_b^\text{max} = 50$Mbps, when $E>3$.
Another highlight of Fig.~\ref{yAR_xMEC_F25_F75} is the remarkable rate improvement of all schemes when more UAVs are employed. More specifically, the average rate has a steep increase for low fronthaul capacities, i.e., in low fronthaul regime, and then experiences more gain while adding more ECs. This underlines the promising gain of the proposed schemes in terms of communication aspects, especially at low fronthaul regime.
Comparing the algorithmic implementations of the three proposed schemes, we note that FCP outperforms PDP and FDP, whereas PDP always outperforms FDP. While the rate differences are low when the number of UAVs is small, FCP reaches increased gaps at high numbers of UAVs, where the presence of a centralized resource management orchestrator is more needed. Interestingly, in high fronthaul regime, PDP has a better gain over FDP and less gap towards FCP, which highlights the superior interference management capability of PDP compared to FDP. In fact, in high fronthaul regime, the interference within the network becomes problematic, which has to be tackled by sophisticated interference management techniques, e.g., FCP and PDP.
In fact, in such regimes, PDP shows considerable gain over FDP and reasonable loss against FCP. Please note that FCP has a relatively higher complexity, run-time (see Sec.~\ref{ssec:ndrt}), and is often difficult to realize in practice, which makes PDP more favorable in future dense networks.\vspace*{-.2cm}
\subsection{Dynamics of the Delay Constraint}
\begin{figure}[!t]
\centering
\vspace*{-.1cm}
\includegraphics[width=2.8in]{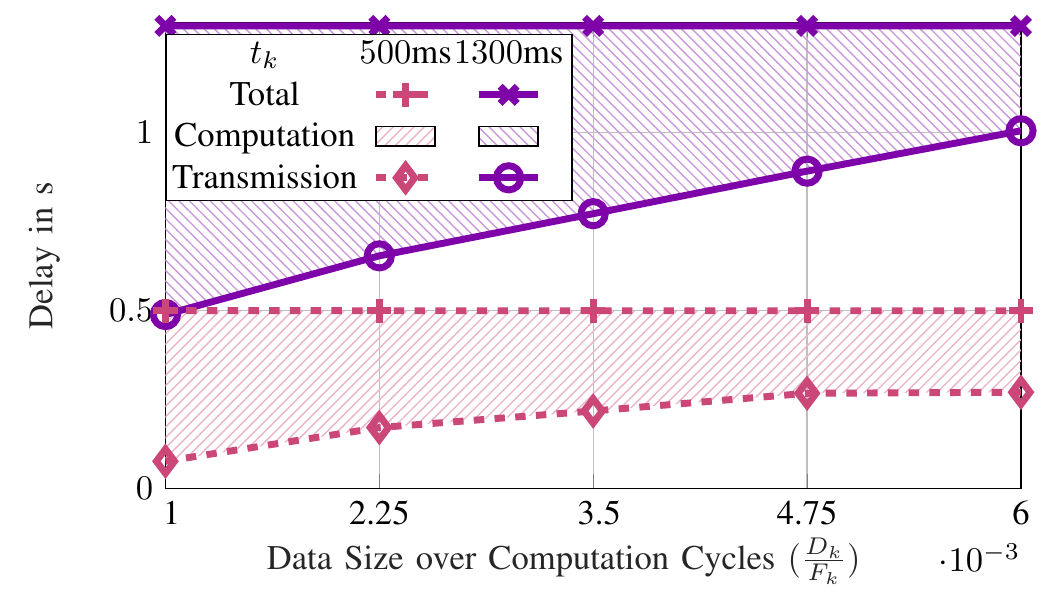}
\vspace*{-.4cm}
\caption{Worst-case delay vs. ${D_k}/{F_k}$.}
\label{yWCD_xDatF}
\vspace*{-.05cm}
\end{figure}

\vspace*{-.1cm}
This part illustrates the numerical dynamics of the delay expression $\Theta_{k}$ \eqref{eq:delaycc}, especially those related to the computation and transmission delays while implementing PDP. Consider a network with varying data size $D_k$, $K=12$, $E = 4$, $L_c=2$, and $P_e^\text{max} = 20$dBm. Fig.~\ref{yWCD_xDatF} shows the worst-case delay, i.e., delay of the user with highest delay, versus the ratio of data size and computation cycles per task $\frac{D_k}{F_k}$ for PDP.
The rationale of such study is that, regardless of the delay maximum value, i.e., $t_k$, Fig.~\ref{yWCD_xDatF} captures the trade-off between communication and computation delay of the considered system model. When $\frac{D_k}{F_k}$ is small, i.e., the data size is small, the worst-case delay mostly consists of the time it takes to compute the task, whereas the transmission delay makes up only a fraction of the total delay. With increasing $\frac{D_k}{F_k}$, the transmission delay becomes more prominent, as more data needs to be transmitted. While $r_k$ is already optimized and no major improvements are possible, the algorithm shifts the resource allocation towards assigning more computation capacity $f_k$ so as to respect the total delay constraint.
Another notable aspect in Fig.~\ref{yWCD_xDatF} is the different ratio of transmission and computation delay for the two considered $t_k$ values. More specifically, at $\frac{D_k}{F_k}=6\cdot10^{-3}$, i.e., the data size is more prominent, the ratio of transmission and computation delay is about $50\%$ for $t_k=500$ms, and about $80\%$ for $t_k=1300$ms, respectively.
One the one hand, PDP is a feasible procedure even in latency-constrained networks, and on the other hand, this result emphasizes the fairness-factor among network participants as delay constraints tighten, since less communication resources are able to be sacrificed from the worst-case user towards the well-connected users.	
As per Fig.~\ref{yWCD_xDatF}, the proposed algorithm's feasibility in terms of strict and loose maximum delay constraints and its versatility in terms of varying ratio of data size and computation cycles are highlighted. 
In other words, the results in Fig.~\ref{yWCD_xDatF} emphasize the trade-off capability of the proposed scheme, as communication and computation resources are dynamically adjusted towards achieving the network management goals. Such valuable results illustrate the numerical prospects of PDP and its promising applicability in the context of future 6G networks.
\subsection{Network Density and Run Time}\label{ssec:ndrt}\vspace*{-.1cm}
We lastly set $P_e^\text{max} = 20$dBm, $E = 4$, and vary the number of users per km$^2$. The runtimes normalized to FDP's runtime with $10.9$ users/km$^2$ are then given in Tab.~\ref{tb:runtime} for the considered protocols. First, we note that PDP achieves comparable runtimes to FDP, which highlights the numerical performance gains of PDP, as it provides better rates and delays in a similar runtime. Focusing on the scalability, both PDP and FDP roughly double (triple) their runtimes at approximately double (triple) the number of users, i.e., $20.3$ ($29.7$) users/km$^2$, whereas FCP takes almost three (four) times longer. 
These results emphasize the gain of PDP and FDP in terms of scalability and runtime advantages, and additionally underline PDP's superiority to FDP in the context of joint communication and computation model adopted in this paper.

\section{Conclusion}\label{sec:con}\vspace{-.1cm}
Future networks necessitate the explicit management of joint communication and computation resources, so as to satisfy the critical requirements of the expected 6G massive deployment. To this end, this paper proposes, and evaluates the benefit of, one particular hybrid central/MEC platform, especially introduced to balance the network resources required for joint computation and communication. The paper particularly focuses on maximizing the weighted sum-rate subject to per-BS and per-UAV power, per-BS fronthaul capacity, per-device maximum computation capacity, and per-user delay constraints, so as to determine the optimal allocated rate, beamforming vectors, and computational capacities. Thanks to $\ell_0$-norm relaxation, SCA, and fractional programming, three different algorithms are proposed to solve the intricate optimization problem. The impacts of network parameters on the rate and delay are then illustrated in the simulation results, which highlight the numerical prospects of the proposed algorithms for enabling joint communication and computation, especially the appreciable improvements of the data processing delays and throughputs as compared to conventional system strategies. At strong interference levels, our proposed distributed algorithm, i.e., PDP, particularly achieves reasonable gains and superior runtime advantages, emphasizing its applicability to future networks applications. \vspace*{-.2cm}
\begin{table}[t]
\vspace*{.2cm}
\renewcommand{\arraystretch}{1.0}
\centering
\begin{tabular}{c c c c c c}
\hline
Users/km$^2$ & $10.9$ & $15.6$ & $20.3$ & $25$ & $29.7$\\\hline
FCP & $1.37$ & $2.19$ & $2.92$ & $3.40$ & $4.26$\\
PDP &  $0.95$ & $1.58$ & $2.21$ & $2.74$ & $3.45$\\
FDP & $1$ & $1.56$ & $2.20$ & $2.69$ & $3.37$\\\hline
\end{tabular}
\vspace*{-.2cm}
\caption{Normalized runtimes.}
\label{tb:runtime}
\vspace*{-.1cm}
\end{table}%


%
%




%

\bibliographystyle{IEEEtran}
\bibliography{bibliography}

\begin{thebibliography}{10}
\providecommand{\url}[1]{#1}
\csname url@samestyle\endcsname
\providecommand{\newblock}{\relax}
\providecommand{\bibinfo}[2]{#2}
\providecommand{\BIBentrySTDinterwordspacing}{\spaceskip=0pt\relax}
\providecommand{\BIBentryALTinterwordstretchfactor}{4}
\providecommand{\BIBentryALTinterwordspacing}{\spaceskip=\fontdimen2\font plus
\BIBentryALTinterwordstretchfactor\fontdimen3\font minus
  \fontdimen4\font\relax}
\providecommand{\BIBforeignlanguage}[2]{{%
\expandafter\ifx\csname l@#1\endcsname\relax
\typeout{** WARNING: IEEEtran.bst: No hyphenation pattern has been}%
\typeout{** loaded for the language `#1'. Using the pattern for}%
\typeout{** the default language instead.}%
\else
\language=\csname l@#1\endcsname
\fi
#2}}
\providecommand{\BIBdecl}{\relax}
\BIBdecl

\bibitem{emr}
\BIBentryALTinterwordspacing
``Ericsson mobility report {November} 2021,'' Ericson, Tech. Rep., Nov. 2020.
  [Online]. Available:
  \url{https://www.ericsson.com/en/mobility-report/reports/november-2021}
\BIBentrySTDinterwordspacing

\bibitem{8961914}
F.~Zhou, R.~Q. Hu, Z.~Li, and Y.~Wang, ``Mobile edge computing in unmanned
  aerial vehicle networks,'' \emph{IEEE Wirel. Commun.}, vol.~27, no.~1, pp.
  140--146, 2020.

\bibitem{8016573}
Y.~Mao, C.~You, J.~Zhang, K.~Huang, and K.~B. Letaief, ``A survey on mobile
  edge computing: The communication perspective,'' \emph{IEEE Commun. Surv.
  Tutor.}, vol.~19, no.~4, pp. 2322--2358, 2017.

\bibitem{Journal}
R.-J. Reifert, A.~A. Ahmad, H.~Dahrouj, A.~Chaaban, A.~Sezgin, T.~Y.
  Al-Naffouri, and M.-S. Alouini, ``Distributed resource management in downlink
  cache-enabled multi-cloud radio access networks,'' \emph{IEEE Trans. Veh.
  Technol.}, pp. 1--16, 2022.

\bibitem{DaiY14}
B.~{Dai} and W.~{Yu}, ``Sparse beamforming and user-centric clustering for
  downlink cloud radio access network,'' \emph{IEEE Access}, vol.~2, pp.
  1326--1339, 2014.

\bibitem{8809879}
X.~Diao, J.~Zheng, Y.~Wu, Y.~Cai, and A.~Anpalagan, ``Joint trajectory design,
  task data, and computing resource allocations for {NOMA}-based and
  {UAV}-assisted mobile edge computing,'' \emph{IEEE Access}, vol.~7, pp.
  117\,448--117\,459, 2019.

\bibitem{8434285}
F.~Zhou, Y.~Wu, R.~Q. Hu, and Y.~Qian, ``Computation rate maximization in
  {UAV}-enabled wireless-powered mobile-edge computing systems,'' \emph{IEEE J.
  Sel. Areas Commun.}, vol.~36, no.~9, pp. 1927--1941, 2018.

\bibitem{8757041}
A.~A. Ahmad, J.~Kakar, R.-J. Reifert, and A.~Sezgin, ``{UAV}-assisted {C-RAN}
  with rate splitting under base station breakdown scenarios,'' in \emph{IEEE
  ICC Workshops}, 2019, pp. 1--6.

\bibitem{8580994}
J.~Kakar, A.~Chaaban, V.~Marojevic, and A.~Sezgin, ``{UAV}-aided multi-way
  communications,'' in \emph{IEEE PIMRC}, 2018, pp. 1169--1173.

\bibitem{8764580}
Z.~Yang, C.~Pan, K.~Wang, and M.~Shikh-Bahaei, ``Energy efficient resource
  allocation in {UAV}-enabled mobile edge computing networks,'' \emph{IEEE
  Trans. Wirel. Commun.}, vol.~18, no.~9, pp. 4576--4589, 2019.

\bibitem{9461747}
W.~Feng, J.~Tang, N.~Zhao, X.~Zhang, X.~Wang, K.-K. Wong, and J.~A. Chambers,
  ``Hybrid beamforming design and resource allocation for {UAV}-aided
  wireless-powered mobile edge computing networks with {NOMA},'' \emph{IEEE J.
  Sel. Areas Commun.}, vol.~39, no.~11, pp. 3271--3286, 2021.

\bibitem{8445936}
X.~Cao, J.~Xu, and R.~Zhang, ``Mobile edge computing for cellular-connected
  {UAV}: Computation offloading and trajectory optimization,'' in \emph{IEEE
  SPAWC}, 2018, pp. 1--5.

\bibitem{9615116}
J.~Xu, K.~Ota, M.~Dong, and H.~Zhou, ``{MCTS}-enhanced hybrid offloading for
  aerial multi-access edge computing,'' \emph{IEEE Wirel. Commun.}, vol.~28,
  no.~5, pp. 82--87, 2021.

\bibitem{8952621}
Q.~Zhang, J.~Chen, L.~Ji, Z.~Feng, Z.~Han, and Z.~Chen, ``Response delay
  optimization in mobile edge computing enabled {UAV} swarm,'' \emph{IEEE
  Trans. Veh. Technol.}, vol.~69, no.~3, pp. 3280--3295, 2020.

\bibitem{8877759}
T.~Zhang, Y.~Xu, J.~Loo, D.~Yang, and L.~Xiao, ``Joint computation and
  communication design for {UAV}-assisted mobile edge computing in {IoT},''
  \emph{IEEE Trans. Ind. Inform.}, vol.~16, no.~8, pp. 5505--5516, 2020.

\bibitem{FP1}
K.~{Shen} and W.~{Yu}, ``Fractional programming for communication systems part
  {I}: Power control and beamforming,'' \emph{IEEE Trans. Signal Process.},
  vol.~66, no.~10, pp. 2616--2630, May 2018.

\bibitem{cvx}
\BIBentryALTinterwordspacing
M.~Grant and S.~Boyd, ``{CVX}: Matlab software for disciplined convex
  programming, version 2.1,'' 2014. [Online]. Available:
  \url{http://cvxr.com/cvx}
\BIBentrySTDinterwordspacing

\bibitem{Lobo1998ApplicationsOS}
M.~Lobo, L.~Vandenberghe, S.~P. Boyd, and H.~Lebret, ``Applications of
  second-order cone programming,'' \emph{Linear Algebra and its Applications},
  vol. 284, pp. 193--228, 1998.

\bibitem{pl}
A.~Al-Hourani, S.~Kandeepan, and S.~Lardner, ``Optimal {LAP} altitude for
  maximum coverage,'' \emph{IEEE Wireless Commun. Lett.}, vol.~3, no.~6, pp.
  569--572, Dec 2014.

\end{thebibliography}

\end{document}